\pdfoutput=1

\documentclass[10pt]{article}

\usepackage{listings}    
\usepackage{graphicx}    
\usepackage{subcaption}  
\usepackage{algorithm}   
\usepackage{algorithmic} 
\usepackage{booktabs}    
\usepackage{hyperref}    
\usepackage[htt]{hyphenat} 
\usepackage{microtype} 
\DisableLigatures{}    

\usepackage{caption}
\usepackage{setspace}
\usepackage{multirow}
\usepackage{enumerate}
\usepackage{pdflscape}
\usepackage{pgfplots} 

\usepackage{xcolor}
\definecolor{codegreen}{rgb}{0.25,0.5,0.35}
\definecolor{codegray}{rgb}{0.5,0.5,0.5}
\definecolor{codepurple}{rgb}{0.6,0,0}
\definecolor{backcolour}{rgb}{0.95,0.95,0.92}
\definecolor{colorstring}{rgb}{0.5,0,0.35}
\definecolor{rltred}{rgb}{0.5,0,0}
\definecolor{rltgreen}{rgb}{0,0.5,0}
\definecolor{rltblue}{rgb}{0,0,0.5}
\definecolor{DarkGreen}{rgb}{0.00,0.60,0.00}
\definecolor{ScarletRed}{rgb}{0.80,0.00,0.00}
\definecolor{blizzardblue}{rgb}{0.67, 0.9, 0.93}
\definecolor{green-yellow}{rgb}{0.68, 1.0, 0.18}
\definecolor{dkgreen}{rgb}{0,0.6,0}
\definecolor{gray}{rgb}{0.5,0.5,0.5}
\definecolor{mauve}{rgb}{0.58,0,0.82}
\definecolor{lightgrey}{rgb}{0.90,0.90,0.90}
\definecolor{grey}{gray}{0.75}
\definecolor{light-gray}{gray}{0.80}

\lstdefinestyle{mystyle}{
    escapechar=©, 
	backgroundcolor=\color{backcolour},
    basicstyle=\footnotesize\ttfamily,
   	identifierstyle=\footnotesize\ttfamily,
	commentstyle=\color{codegreen},
	keywordstyle=\color{colorstring}\bfseries,
	numberstyle=\ttfamily\color{codegray},
	stringstyle=\ttfamily\color{DarkGreen},
	breakatwhitespace=false,
	breaklines=true,
	captionpos=b,
	keepspaces=true,
	numbers=left, 
	numbersep=2pt,
	showspaces=false,
	showstringspaces=false,
	showtabs=false,
	tabsize=2
}
\usepackage{xspace}
\newcommand{\evo}{{\sc EvoMaster}\xspace}

\usepackage{boxedminipage}
\newenvironment{result}%
{\smallskip
	\noindent
	\let\emph=\textbf
	\begin{boxedminipage}{\columnwidth}\begin{center}\em}%
		{\end{center}\end{boxedminipage}%
}

\usepackage{ifthen}
\newboolean{showcomments}
\setboolean{showcomments}{true} 

\ifthenelse{\boolean{showcomments}}{
	\newcommand{\nbc}[3]{
		{\colorbox{#3}{\bfseries\sffamily\scriptsize\textcolor{white}{#1}}}
		{\textcolor{#3}{\sf\small$\langle$\textit{#2}$\rangle$}}}
	
}{
	\newcommand{\nbc}[3]{}

}

\newcommand{\totalLinkedIn}{$39$\xspace}

\usepackage{mathtools} 
\usepackage{subcaption}  
\usepackage[most]{tcolorbox} 
\usepackage{soul} 
\usepackage{makecell} 

\usepackage{authblk}                
\usepackage[square,numbers]{natbib} 
\usepackage{amsmath}                

\newcommand{\NameAssist}{{\sc NameAssist}\xspace}
\newcommand{\DeepTCEnhancer}{{\sc DeepTC-Enhancer}\xspace}
\newcommand{\codeToSeq}{{\sc code2seq}\xspace}
\newcommand{\NATIC}{{\sc NATIC}\xspace}
\newcommand{\EvoSuite}{{\sc EvoSuite}\xspace}
\newcommand{\Sapienz}{{\sc Sapienz}\xspace}
\newcommand{\RESTAssured}{{\sc REST-Assured}\xspace}
\newcommand{\GitHub}{{\sc GitHub}\xspace}
\newcommand{\LinkedIn}{{\sc LinkedIn}\xspace}

\newcommand{\numbered}{{\sc testNumber}\xspace}
\newcommand{\evosimple}{{\sc testMethodOnEndpointResult}\xspace}
\newcommand{\evoquery}{{\sc testMethodOnQualifiedEndpointWithQueryResult}\xspace}
\newcommand{\evocondition}{{\sc testMethodOnQualifiedEndpointWithQueryConditionsResult}\xspace}
\newcommand{\evosimpleshort}{{\sc testResult}\xspace}
\newcommand{\evoqueryshort}{{\sc testQuery}\xspace}
\newcommand{\evoconditionshort}{{\sc testCondition}\xspace}

\usepackage{geometry}
\title{
Generating REST API Tests With Descriptive Names
}

\author[1,3]{Philip Garrett}

\author[1,2,3]{Juan P. Galeotti}

\author[3,4]{Andrea Arcuri}

\author[5]{Alexander Poth}

\author[5]{Olsi Rrjolli}

\affil[1]{Computer Science Dept., School of Exact and Natural Sciences, University of Buenos Aires, Buenos Aires, Argentina}

\affil[2]{Institute of Computer Science Research, CONICET, Buenos Aires, Argentina}

\affil[3]{School of Economics, Innovation and Technology, Kristiania University of Applied Sciences, Oslo, Norway}

\affil[4]{Department of Computer Science, Oslo Metropolitan University, Oslo, Norway}

\affil[5]{Volkswagen AG, Germany}

\date{}

\begin{document}

\maketitle

\begin{abstract}
Automated test generation has become a key technique for ensuring software quality, particularly in modern API-based architectures.
However, automatically generated test cases are typically assigned non-descriptive names (e.g., \texttt{test0}, \texttt{test1}), which reduces their readability and hinders their usefulness during comprehension and maintenance.
While prior research has explored naming strategies for unit tests, little attention has been given to system-level REST API tests, where meaningful names should capture endpoints, parameters, and expected outcomes.

In this work, we present three novel deterministic techniques to generate test names.
We then compare eight techniques in total for generating descriptive names for REST API tests automatically produced by the fuzzer \evo, using 10 test cases generated for 9 different open-source APIs.
The eight techniques include  rule-based heuristics and large language model (LLM)-based approaches.
Their effectiveness was empirically evaluated through two surveys (involving up to \totalLinkedIn people recruited via LinkedIn).
Our results show that a rule-based approach achieves the highest clarity ratings among deterministic methods, performs on par with state-of-the-art LLM-based models such as \emph{Gemini} and \emph{GPT-4o}, and significantly outperforms \emph{GPT-3.5}.

To further evaluate the practical impact of our results, an industrial case study was carried out with practitioners who actively use \evo at Volkswagen AG.
A developer questionnaire was then carried out based on the use of \evo on four different APIs by four different users, for a total of 74 evaluated test cases.
Feedback from practitioners further confirms that descriptive names produced by this approach improve test suite readability.

These findings highlight that lightweight, deterministic techniques can serve as effective alternatives to computationally expensive and security-sensitive LLM-based approaches for automated system-level test naming, providing a practical step toward more developer-friendly API test generation.

\end{abstract}

{\bf Keywords}: REST API, Test Case Naming, Search-based Software Testing, Large Language Models, Automated Test Case Generation, Human Study

\section{Introduction}

Software testing is a fundamental part of software development. It is essential for verifying program behavior and ensuring the quality of software systems.
Writing test cases manually can be a tedious, costly, and potentially incomplete task, as certain conditions might require inputs that are difficult to reproduce manually or may not even be anticipated by developers~\cite{Daka_2014}.
This has led to significant interest in automated test generation techniques~\cite{Saswat_2013}.

In modern software architectures, particularly those based on microservices, APIs (such as REST~\cite{fielding2000architectural}, GraphQL~\cite{GraphQLFoundation}, and RPC~\cite{gRPC,thrift}) have become popular for exposing services and defining communication interfaces.
Several tools have been proposed for automatically generating test cases for these types of APIs~\cite{arcuri2019restful,martinLopez2021Restest,restlerICSE2019,viglianisi2020resttestgen,laranjeiro2021black,hatfield2022deriving,wu2022icse,benac2014jsongen}.
These techniques include both black and white-box approaches to synthesize the generated test cases.

However, a common challenge with automatically generated tests is that they often receive non-descriptive names, such as ``\texttt{test0}'' or ``\texttt{test1}''.
Unlike manually written tests, automatically generated tests may not represent real scenarios and might primarily aim to cover code, making intuitive naming difficult.
Such generic names fail to provide the benefits that descriptive names offer.
Descriptive test names are highly valuable as they help developers to understand the goal and scenario of a test case, supporting navigation through test suites~\cite{daka2017generating}.
When trying to understand code, unit tests can serve as usage examples, and good names simplify this.
During maintenance, test names help identify undesired side-effects and facilitate updating tests when code changes.
Providing tests with good names simplifies these tasks, which is crucial given the substantial costs of software maintenance~\cite{Lientz_1980}.

While previous research has explored naming for unit tests~\cite{allmanis2015,Yonai2019,roy2020,Biagiola_2025,zhang2016,daka2017generating,nijkamp2021,Wu_2023}, automatically generating system-level API tests present unique challenges and opportunities for test naming.
REST API tests exercise endpoints with specific HTTP methods, parameters, and expect particular response types or status codes.
Simply applying unit test naming strategies might not fully capture the behavior and purpose of an API test.

This paper focuses specifically on the problem of generating descriptive names for automatically generated REST API tests.
We propose novel deterministic approaches to provide meaningful names.
We also evaluate and compare different existing techniques for generating test case names, considering both deterministic and probabilistic approaches like the use of LLMs.
Our goal is to determine which strategy best fits the needs of automatic test generation tools, ensuring that generated test names remain descriptive, unique, and clearly connected to the exercised API endpoints.
Through this evaluation, we aim to lay the groundwork for more effective and developer-friendly automated test generation for REST APIs

As test readability is a metric that requires  \emph{human} evaluation, we carried out empirical studies to assess it.
First, we recruited \totalLinkedIn participants from LinkedIn.
Each participant was tasked to evaluate the readability and usefulness of names of up to 10 different test cases generated for different open-source projects using the state-of-the-art fuzzer \evo~\cite{arcuri2025tool}.
Each presented test was given different options for their names, and the human participants were asked to rank those options.
Participants were not told which strategies were used to create those names.

Once this study was completed, the best identified strategy (which is \evoconditionshort) was integrated in \evo.
This strategy is now the default in \evo since version $4.0.0$~\cite{zenodo400evomaster}.
However, a potential limitation of our study is that practitioners evaluated readability on isolated test cases for open-source APIs they have no involvement with.
To improve realism, and see the actual consequences of using these techniques in practice, we carried out as well an industrial study.
As \evo is a state-of-the-art tool used in several enterprises around the world, we asked test engineers at Volkswagen AG (a German car manufacturer) to apply \evo on four of their APIs they have been using \evo on.
We prepared a questionnaire to get insight on how this new naming strategy impacted \emph{all} the test cases (74 in total) generated for these four APIs compared to the previous default naming strategy (which was just a numeric counter like \texttt{test0()} and \texttt{test1()}).
Results were positive, where the new naming strategy  was evaluated as a significant improvement.

Besides test naming, another important aspect for test readability is how the \emph{test cases} are organized within a \emph{test suite} file.
We have developed a strategy to organize test cases by providing meaningful sorting based on REST' semantics.
As the evaluation of this strategy requires to analyze whole test suite files (possibly with hundreds of lines of code), it is much more time consuming and demanding than looking at test names of test cases in isolation.
As such, no controlled empirical experiment was carried out for this new feature.
We rather based its evaluation on the feedback provided in the developer questionnaire done for the four API used as industrial case study at Volkswagen AG.

The remainder of this article is organized as follows.
Section~\ref{sec:background} provides the necessary background on automated test generation and the challenges of naming test cases.
Section~\ref{sec:related_work} reviews related work on test name generation, including probabilistic approaches and LLM-based approaches.
Section~\ref{sec:naming_convention} describes the different approaches considered in our empirical study, including rule-based and LLM-based approaches.
Section~\ref{sec:evaluation} presents our empirical evaluation, including the experimental setup, both survey and questionnaire results, and statistical analyses.
Section~\ref{sec:threats_to_validity} discusses threats to validity and potential limitations of our study.
Finally, Section~\ref{sec:conclusions} summarizes our main findings, draws conclusions, and provides directions for future work.

\section{Background}
\label{sec:background}
\subsection{RESTful APIs}

RESTful APIs~\cite{fielding2000architectural} are web services that use HTTP(S) calls to uniformly expose and operate on resources  over the internet.
An HTTP(S) call performs a given action (e.g., \texttt{GET} and \texttt{POST}) on a specific endpoint.
To help interoperability, RESTful API endpoints are generally published using a uniform machine-readable specification such as OpenAPI~\cite{openapispec}.
As an example, an OpenAPI schema for a given RESTful API could specify that a given endpoint is accessible by satisfying a given pattern such as ``\texttt{/users/\{userId\}/orders/\{productId\}}''.
A call targeted towards this endpoint must provide actual values for the path parameters (namely, \texttt{userId} and \texttt{productId}).
The following path ``\texttt{/users/42/orders/1234}'' conforms to the given endpoint pattern by providing the actual values and matching the expected static/fixed segment values (in the previous case, ``\texttt{users}'' and ``\texttt{orders}'').
Query parameters could also be appended to the path by adding the ``\texttt{?}'' symbol.
As an example, in the following path  ``\texttt{/users/42/orders/1234?includeDetails=true\&currency=EUR}'', we observe the previously base path plus query parameters ``\texttt{includeDetails}'' and ``\texttt{currency}'' set to ``\texttt{true}'' and ``\texttt{EUR}'',  respectively.

Once the HTTP(S) call is performed, the server returns a response.
This response contains a \emph{status code} (a 3-digit number indicating the result of the request, such as \texttt{200}, \texttt{404}, etc.), and optionally might include a \emph{response body} that could be encoded in a specific format (such as JSON~\cite{jsonspecwright}, HTML~\cite{fielding1999hypertext}, XML, etc.).
As an example, the JSON object in Figure~\ref{fig:json-response-body} is returned from the server in the response body after a \texttt{GET} call to endpoint   ``\texttt{/users/42/orders/1234}'' with query parameters
``\texttt{includeDetails=true}'' and ``\texttt{currency=EUR}''.

REST API test cases are generally defined as a sequence of HTTP(S) calls.
They might also include assertions to check expected conditions on the result of each of such calls.
\RESTAssured~\cite{RestAssured} is a Java library for testing and validating RESTful APIs.
It allows to write declarative and readable REST API test cases in combination with testing frameworks such as JUNIT~\cite{JUNIT}.
In Figure~\ref{fig:rest-assured-test} it is shown a test case that performs an HTTP \texttt{GET}  request to a server located in \texttt{baseUrlOfSut}.
Following this request,  the test case checks that the server response returns a \texttt{200} status code, and the response also includes a response body in the JSON format.
Finally, the test case checks that the returned JSON object has a field named \texttt{currency} whose value is ``\texttt{EUR}''.
If any of these checks are not satisfied, the test case fails.

\begin{figure}[!t]
\centering
\begin{lstlisting}
@Test
public void test1() {
    given().
        baseUri(baseUrlOfSut).
    when().
        get("/users/42/orders/1234?includeItems=true&currency=EUR").
    then().
        statusCode(200).
        assertThat().
        contentType("application/json").
        body("currency", containsString("EUR"));
}
\end{lstlisting}
\caption{\RESTAssured test checking currency in the response}
\label{fig:rest-assured-test}
\end{figure}

\begin{figure}[!t]
\centering
\begin{lstlisting}
{
  "orderId": "1234",
  "userId": "42",
  "totalAmount": 99.99,
  "currency": "EUR",
  "items": [
    {
      "itemId": "A1",
      "name": "Product X",
      "price": 50.00
    },
    {
      "itemId": "B2",
      "name": "Product Y",
      "price": 49.99
    }
  ],
  "status": "completed"
}
\end{lstlisting}
\caption{The JSON response body obtained from executing the call in \texttt{test1}}
\label{fig:json-response-body}
\end{figure}

\subsection{EvoMaster}

Over the years, search-based software engineering has emerged as a highly effective approach to tackle a broad spectrum of software engineering challenges~\cite{harman2012search}, particularly in the realm of software testing~\cite{ABHP09}.
Notable tools such as \EvoSuite~\cite{fraser2011evosuite}, for Java unit testing, and \Sapienz~\cite{mao2016sapienz}, for Android testing, exemplify the success of these techniques.

Software testing can be cast as an optimization problem, maximizing both code coverage and fault detection in the generated test suites.
Once a fitness function is established for the testing problem, a search algorithm can explore the solution space, which in this case represents all possible test cases.
Several types of search algorithms exist, with Genetic Algorithms~\cite{PHP99} being among the most popular.
Specialized search algorithms (such as Whole Test Suite~\cite{GoA_TSE12}, Many-Objective Sorting Algorithm (MOSA)~\cite{dynamosa2017} and the Many Independent Objective Algorithm (MIO)~\cite{arcuri2018test}) have been developed for the specific problem of generating test suites (in contrast to solely generating test cases).

\evo~\cite{arcuri2021evomaster,arcuri2018evomaster} is an open-source tool designed to generate system-level test cases for REST APIs~\cite{arcuri2019restful}.
It can generate test cases without internal knowledge of the system being tested (i.e., black-box mode) or by analysing internal runtime of the application (white-box).
When the white-box approach is chosen, \evo can leverage evolutionary algorithms, such as the MIO~\cite{arcuri2018test} and MOSA~\cite{dynamosa2017}, to achieve higher coverage and fault detection.
However, this approach is only supported for REST APIs that are compiled into JVM bytecode (e.g., Java and Kotlin).
In this setting, \evo is composed of two main components:
(1) the \emph{core}, which handles the command-line interface, search algorithms, and test case generation;
and (2) the \emph{controller} library, which users employ to write configuration classes that instruct \evo on how to start, reset, and stop the system under test (SUT).
The controller library also automatically instruments the SUT at startup to collect runtime heuristics, such as the aforementioned branch distance.

\evo produces  self-contained test cases that can be written in multiple languages, including Java, Kotlin, Javascript, and Python~\cite{arcuri2025widening}.
Generated tests for white-box testing of JVM-based APIs interact with a \emph{controller}, which manages the lifecycle of the SUT by automatically starting, stopping, and resetting it through  \texttt{@Before} and \texttt{@After} method hooks.
These test cases can be run directly within an IDE (e.g., IntelliJ) or incorporated into build automation tools as Maven or Gradle.
Each test generated in Java or Kotlin consists of a sequence of HTTP requests with the \RESTAssured~\cite{RestAssured} library.
Figure~\ref{fig:evomaster-generated-test} shows a test case generated by \evo for the \emph{news} REST API~\cite{icst2023emb}.
In this test case, \evo not only performs an action on a RESTful API endpoint (i.e., the \texttt{GET} HTTP request on \texttt{/news}), but also injects data directly into the underlying SQL database
by means of a specific Domain Specific Language (DSL) for such actions.
In particular, the invokation to the \texttt{execInsertionsIntoDatabase} method with the data that will be inserted into a \texttt{NEWS\_ENTITY} table in the SQL database.

\evo can also generate test cases for injecting data for both SQL (e.g., Postgres, MySQL, etc.) and NoSQL databases (e.g., Mongo~\cite{MongoDB}), and mock external microservices by means of the WireMock library~\cite{seran2025handling}.
\evo is open-source on GitHub and can be accessed at \url{https://www.evomaster.org}.

\begin{figure}[!t]
\centering
\begin{lstlisting}
@Test(timeout = 60000)
public void test_3() throws Exception {
    List<InsertionDto> insertions = sql().insertInto("NEWS_ENTITY", 138L)
            .d("ID", "0")
            .d("AUTHOR_ID", "\"_EM_8387_XYZ_\"")
            .d("COUNTRY", "\"UHrRU\"")
            .d("CREATION_TIME", "\"1932-05-10 00:18:37\"")
            .d("TEXT", "\"BTm9EKaK_\"")
        .dtos();
    controller.execInsertionsIntoDatabase(insertions);

    given().accept("application/vnd.tsdes.news+json;charset=UTF-8;version=2")
            .header("x-EMextraHeader123", "")
            .get(baseUrlOfSut + "/news?authorId=Z7R6YC7R9Sn_HJ&country=")
            .then()
            .statusCode(200)
            .assertThat()
            .contentType("application/vnd.tsdes.news+json")
            .body("size()", equalTo(0));
}
\end{lstlisting}
\caption{A test case automatically generated by \evo for the \emph{news} REST API.}
\label{fig:evomaster-generated-test}
\end{figure}

\subsection{Test naming conventions}

In the domain of computer science, a \emph{naming convention} is a specific set of rules describing the selection of character sequences for identifiers.
These identifiers serve to uniquely denote entities in the programming language domain such as variables, data types, functions, and other programmatic entities within both source code and associated documentation~\cite{Feitelson_2022}.
\emph{Method naming conventions} refer to the subset of those rules that apply on how methods are identified in a given object-oriented programming language~\cite{Host_2009}.
As test cases are written as methods, \emph{test naming convention} describes which rules should be applied to identify a test case within a given test class~\cite{Robillard_2025}.

In unit testing, a test naming convention includes semantic information about the unit test case~\cite{Wang_2025}.
As an example, the unit test naming convention \texttt{testMethod\_Result} identifies which is the method under test (also known as the \emph{focal method}~\cite{Ghafari_2015}) using the \texttt{Method} placeholder, while the expected result of the execution of the test case is described in the \texttt{Result} placeholder.
The test naming convention might also include syntactic features to improve both readability and understandibility of the test case.
In the previous example, a test prefix marker (i.e., the word ``\texttt{test}'') is included, as well as a separator (i.e, the symbol ``\texttt{\_}'') to delimit the fragments of the semantic information.
An instance of the above test naming convention could be ``\texttt{testStartClientSession\_success}''.
Other fragments of semantic information a test naming convention could include are the class under test, the method under test, the input state related to the method under test, the result of the method execution, or a general description of the specific scenario that is being exercised~\cite{Wang_2025}.

\section{Related Work}
\label{sec:related_work}

\subsection{Rule-Based/Coverage-Based Approaches}


Recent approaches were proposed to analyze existing human written test names, extracting specific test information~\cite{wu_2020, Wang_2025}, assess quality~\cite{peruma_2021}, or generate test templates~\cite{Zhang_2015}.
Techniques categorized as \emph{rule-based} or \emph{coverage-based}~\cite{zhang2016,daka2017generating,nijkamp2021,Wu_2023} share two key features (i) they rely on an algorithmic procedure grounded in predefined templates or rules for test naming, and (ii) they extract information from the generated test cases (such as exercised methods or covered goals).

\NameAssist~\cite{zhang2016} is a deterministic rule-based approach that automatically generates descriptive test names.
It extracts three key components from the test body: the method under test, the expected outcome (i.e., the assertion) and the test scenario (i.e., parameters and context of the exercised method).
This information is then translated into a descriptive test name using customizable templates.
However, \NameAssist is specifically designed for unit tests containing  a single assertion, and is therefore not directly applicable to REST API testing.

Daka et al.~\cite{daka2017generating} introduce an algorithm to generate test names based on coverage goals (e.g., methods, exceptions thrown, and partitions on input and output values).
Their approach prioritizes coverage goals that are exclusively exercised by each test case when generating names.
The class under test is considered for verifying whether a given method changes the state of an instance (i.e., distinguishing pure from impure methods).
Compared to \NameAssist~\cite{zhang2016}, the approach introduced by Daka et al. can handle tests with multiple assertions, but it remains heavily tailored to unit testing and therefore not applicable in the context of REST API testing.

\NATIC~\cite{nijkamp2021} extends the work of Daka et al.~\cite{daka2017generating} with a focus on test amplification (i.e., synthesizing new test cases from existing ones).
It generates test names based on the methods additionally covered by the amplified test when compared to the original test suite.
Since the approach relies on existing \emph{declarative} test names in the test suite to be augmented, \NATIC is not suitable for automatically naming test suites that are fully generated.

Wu and Clause~\cite{Wu_2023} present an approach that extracts attributes making a test case \emph{unique} within the context of a test suite.
However the uniqueness identified by their technique is defined in terms of the underlying implementation, rather than the descriptive API-level semantics relevant to REST API testing (e.g., a returning 200 status code).
Furthermore, the authors acknowledge as a limitation that their technique is not well-suited for \emph{``different programming languages and technologies (i.e., ASP.NET, REST)''}~\cite{Wu_2023}.

\subsection{Probabilistic/Machine Learning Approaches}
\label{sec:mlapproaches}

While rule-based and coverage-based techniques rely on predefined templates or coverage goals, some approaches have explored the use of machine learning and neural models for test naming~\cite{allmanis2015,Yonai2019,roy2020,Biagiola_2025}.
These approaches aim to learn naming patterns directly from source code or the test suites, thereby avoiding previously defined rules.
In this context, different strategies have been proposed, ranging from probabilistic language models trained on source code, to graph-based neural encodings, to specialized deep learning architectures for test case naming.

Allmanis et al.~\cite{allmanis2015} present a neural probabilistic language model specifically designed for suggesting accurate method and class names.
The input language of this model is the source code of a given method.
Such model is trained from the method's bodies from the whole project, and a variant is also proposed to allow neologisms (i.e.,  a name that has not appeared in the training corpus).
When it is applied to naming a given test case, the language model heavily relies on variable identifiers used in the test case, which usually are not descriptive in generated test cases.
Yonai et al.~\cite{Yonai2019} explicitly use graph embedding techniques to encode call graphs for the different methods in the program.
This encoding benefits from a clear purpose or scenario, which might be lacking in the context of automatically generated test cases~\cite{daka2017generating}.

In contrast to~\cite{allmanis2015,Yonai2019}, \DeepTCEnhancer~\cite{roy2020} relies on a deep learning model specifically aimed at automatically generated unit test cases.
It suggests test case names using an adapted deep learning model proposed in \codeToSeq~\cite{alon2019}.
The adapted model is trained with a dataset of engineering software projects by Muniah et al.~\cite{Munaiah2017} containing 678,860 unit test cases.
System-level REST API test cases operate at different abstraction layer, focusing on sequences of HTTP requests and response, endpoints, and status codes, rather than directly invoking methods on a class under test.

Biagiola et al.~\cite{Biagiola_2025} propose to improve the readability of automatically generated test cases (which includes test renaming) using available large language models.
Their approach engineers a prompt that includes focal information of the class under test (e.g., name, attributes and methods), the targeted test case and the method's implementations of the class' methods that are called from the test case.

\section{Test Naming Approaches for \RESTAssured test cases}
\label{sec:naming_convention}

\subsection{Rule-Based Approaches}
\label{sec:rule-based}

In this section we propose three rule-based test naming approaches for \RESTAssured test cases.
Each approach incorporates more semantic information from the test case, including query parameters and their specific conditions.
We will explain these test naming approaches by means of the \RESTAssured test case shown in Figure~\ref{fig:rest-assured-test}.

Furthermore, to avoid the generation of too long names,
we enabled a configurable mechanism to limit the length of a test case name to a maximum of $N$ characters (with default value $N=80$).
If a generated name is longer than the limit $N$, then it becomes truncated.

\subsubsection{ \evosimple}

\begin{figure}[!t]
\begin{flushleft}
\begin{tabbing}
\hspace{3cm} \= \kill
\textit{\textless testName\textgreater} \> ::= \texttt{test}\_\textit{\textless index\textgreater}\_\textit{\textless httpMethod\textgreater}\texttt{On}\textit{\textless nameQualifier\textgreater} \\
                         \> \quad  \textit{\textless resultPart\textgreater} \\
                         \> \quad [ \texttt{Using}\textit{\textless mechanism\textgreater} ] \\[1ex]

\textit{\textless resultPart\textgreater} \> ::= \texttt{Returns}\textit{\textless expectedResult\textgreater} \\
                 \> \quad | \quad \textit{\textless faultLabel\textgreater} \\
                 \> \quad | \quad \texttt{ShowsFaults\_}\textit{\textless faultCodes\textgreater} \\[1ex]

\textit{\textless expectedResult\textgreater} \> ::= \textit{\textless statusCode\textgreater} \\
		\> \quad | \quad \texttt{Empty} \\
                 \> \quad | \quad \texttt{EmptyList} \\
                 \> \quad | \quad \textit{\textless number\textgreater} \texttt{element} \\
                 \> \quad | \quad \textit{\textless number\textgreater} \texttt{elements} \\
                 \> \quad | \quad \texttt{EmptyObject} \\
                 \> \quad | \quad \texttt{Object} \\
                 \> \quad | \quad \texttt{String} \\
                 \> \quad | \quad \texttt{Content}
\end{tabbing}
\end{flushleft}
\caption{The BNF grammar for the \evosimple test naming (abbreviated as \evosimpleshort).}
\label{fig:test-naming-evosimple}
\end{figure}

\begin{table}[!t]
\caption{Rules for non-terminal expansion in the \evosimple test naming.}
\begin{center}
\begin{tabular}{l|p{10cm}}
Non-terminals & Description \\
\hline
\textit{\textless httpMethod\textgreater} & The HTTP method used in the request (\texttt{GET}, \texttt{POST}, \texttt{PUT}, \texttt{DELETE}, etc.). \\
\textit{\textless nameQualifier\textgreater} & The primary resource (e.g., \texttt{Refunds}) of the endpoint under test. \\
\textit{\textless statusCode\textgreater} & Describes the expected HTTP status (e.g., \texttt{200}, \texttt{400}, \texttt{404}).\\
\textit{\textless faultLabel\textgreater}& Short textual label for a given fault condition (e.g., \texttt{causes500\_internalServerError}, \texttt{returnsSchemaInvalidResponse}, etc.).\\
\textit{\textless faultCodes\textgreater} & An underscore (i.e.,``\_'') separated list of the expected fault codes (e.g., \texttt{ShowsFaults\_401\_403}). \\
\textit{\textless mechanism\textgreater} & Describes if the test case uses a mocking/stubbing infrastructure, or directly inserts data into a backend database (e.g., \texttt{Mongo}, \texttt{WireMock}, etc.).
\end{tabular}
\end{center}
\label{tab:evosimple-components}
\end{table}

\begin{table}[!t]
\caption{Values for the \textit{\textless expectedResult\textgreater} component as specified by the \RESTAssured assertions on the response body.}
 \begin{center}
 \begin{tabular}{p{6cm}|p{4cm}|c}
 \RESTAssured assertions & Description & \textit{\textless expectedResult\textgreater} \\
\hline
\texttt{.body(isEmptyOrNullString())} or \texttt{.body(equalTo(""))} & null or blank   & \texttt{Empty} \\ [3pt]

\texttt{.body("size()",equalTo(0))} & empty JSON array 		& \texttt{EmptyList} \\ [3pt]

\texttt{.body("size()",equalTo(N))} & non-empty JSON array &  \texttt{Nelement}[s]  \\ [3pt]

\texttt{.body(equalTo("\{\}"))} & JSON object with no fields 		& \texttt{EmptyObject}  \\ [3pt]

\texttt{.body(F,...))} s.t. F $\neq$ ``\texttt{size()}'' & JSON object with some field & \texttt{Object} 	\\[3pt]

\texttt{.body(equalTo(S)))} s.t. S $\neq$ ``'' or ``\{\}''& body is a JSON non-blank textual & \texttt{String} 	 \\[3pt]

Otherwise & none of the previous conditions apply & \texttt{Content} \\
 \end{tabular}
 \end{center}
\label{tab:expected-result}
 \end{table}

As previously stated, the \RESTAssured test cases are composed by a sequence of HTTP calls.
The first rule-based test naming approach we will describe takes into account the HTTP verb/method in the last HTTP call.
Additionally, it will consider not only such method, but also semantic information on the endpoint on which the method is targeted, and information regarding the expected outcome of such call.
The reason to only focus on the last HTTP call is twofold:
(i) previous calls are supposed to be done just to setup the state of the API (e.g., creating data with specific values) to be able to test the last call in the sequence;
(ii) adding information from each call to the final test name might make this latter too long (especially when several calls appear in a test).

Figure~\ref{fig:test-naming-evosimple} presents the grammar of the \evosimple test naming approach (abbreviated as \evosimpleshort) in Backus-Naur Form~\cite{ASU86}.
Since the test case in Figure~\ref{fig:rest-assured-test} contains a unique HTTP call, this call is the one considered by the naming approach to extract the information require to name the test case.
Table~\ref{fig:test-naming-evosimple} details how non-terminals should be expanded according to the test structure.
As an example, the \textit{\textless httpMethod\textgreater}  non-terminal expands to ``\texttt{get}'' as this is the verb used in the last HTTP call in the test case.
Next, as the accessed path is ``\texttt{/users/42/orders/1234}'', the primary resource in this path (i.e., the \textit{\textless nameQualifier\textgreater} non-terminal) is ``\texttt{orders}''.
Notice that, path ``\texttt{/}'' contains no \textit{\textless nameQualifier\textgreater}.
In such scenario, the naming approach defaults to the ``\texttt{root}'' identifier.

Since the HTTP call is a \texttt{GET} call and the expected response status code is the \texttt{200} value, the \textit{\textless expectedResult\textgreater} non-terminal is expanded using the rules described in Table~\ref{tab:expected-result}.
Otherwise, the status code's response is added (e.g., ``\texttt{Returns400}'').
For the test case shown in Figure~\ref{fig:rest-assured-test}, since there is at least one assertion checking the value of an JSON object field (i.e., \texttt{body("currency",...)}), the \textit{\textless expectedResult\textgreater} non-terminal expands to ``\texttt{ReturnsObject}''.

Consequently, the test case name for the test case in Figure~\ref{fig:rest-assured-test} produced with the \evosimpleshort approach will be:
\begin{center}
``\texttt{test\_0\_getOnOrdersReturnsObject}''
\end{center}

Within a given test class, no test name is repeated.
This rule is enforced by the compiler.
Unfortunately, this cannot be enforced by the test naming convention since it is possible to have two test cases that are different but still lead to the same test name components, and consequently, to the same test name definition.
As an example, let us now consider the test case shown in Figure~\ref{fig:rest-assured-test-2}.
As the test name components match those in Figure~\ref{fig:rest-assured-test}, both test cases will have the same test case name.
In order to avoid test name duplication (and consequently, a failure during compilation), an \textit{\textless index\textgreater} numeric component is included to guarantee uniqueness among test cases.
If both test cases where present in the same test class, the approach will choose ``\texttt{test\_0\_getOn...}'' and ``\texttt{test\_1\_getOn...}'' depending on the appearance order within the test class.
Notice that the indexes are added as prefixes instead of suffixes.
This is due to the fact that this ensures ordering, and allows a smoother user experience while locating a failing test case.

\subsubsection{ \evoquery}

\begin{figure}[!t]
\begin{flushleft}
\begin{tabbing}
\hspace{3cm} \= \kill
\textit{\textless testName\textgreater} \> ::= \texttt{test}\_\textit{\textless index\textgreater}\_\textit{\textless httpMethod\textgreater}\texttt{On}[\textit{\textless parentQualifier\textgreater}]\textit{\textless nameQualifier\textgreater} \\
                         \> \quad [ \texttt{WithQueryParams} ] \\
                         \> \quad  \textit{\textless resultPart\textgreater} \\
                         \> \quad [ \texttt{Using}\textit{\textless mechanism\textgreater} ] \\[1ex]
\end{tabbing}
\end{flushleft}
\caption{The BNF grammar for the \evoquery test naming approach (abbreviated as \evoqueryshort)}
\label{fig:test-naming-evoquery}
\end{figure}

It is worth noting that an endpoint could not also refer to a single qualifier but also contain an entire hierarchy of qualifiers.
As an example, the accessed path ``\texttt{/users/42/orders/1234}'', contains not only a primary resource  (i.e., ``\texttt{orders}''), but also a parent resource: ``\texttt{users}''.
The \evoquery test naming approach (abbreviated as \evoqueryshort) extends the approach previously introduced with two additional features: the parent name qualifier (if it exists), and the token ``\texttt{WithQueryParms}'' in case the HTTP call includes query parameters.
Figure~\ref{fig:test-naming-evoquery} presents the grammar for this test naming approach.
Following the test case in Figure~\ref{fig:rest-assured-test}, the endpoint contains the parent qualifier ``\texttt{users}'' on the HTTP path.
What is more, the HTTP call also includes the query parameters \texttt{includeItems=true} and \texttt{currency=EUR}.
Therefore, the resulting test name produced by this approach will be:
\begin{center}
``\texttt{test\_0\_getOn\underline{Users}Orders\underline{WithQueryParams}ReturnsObject}''
\end{center}

This test naming approach is an extension of  the \evosimpleshort test naming.
In order to highlight how these names differ, we underline the added text between the two test names.

\subsubsection{ \evocondition}

\begin{figure}[!t]
\begin{flushleft}
\begin{tabbing}
\hspace{3cm} \= \kill
\textit{\textless testName\textgreater} \> ::= \texttt{test}\_\textit{\textless index\textgreater}\_\textit{\textless httpMethod\textgreater}\texttt{On}[\textit{\textless parentQualifier\textgreater}]\textit{\textless nameQualifier\textgreater} \\
                         \> \quad [ \texttt{WithQueryParams}\textit{\textless queryConditions\textgreater} ] \\
                         \> \quad \textit{\textless resultPart\textgreater} \\
                         \> \quad [ \texttt{Using}\textit{\textless mechanism\textgreater} ] \\[1ex]
\end{tabbing}
\end{flushleft}
\caption{The BNF grammar for the \evocondition test naming approach (abbreviated as \evoconditionshort).}
\label{fig:test-naming-grammar}
\end{figure}

Specifying the conditions that hold on the query parameters can enrich the test case name.
This feature can help developers understanding the specific scenario that the test case is exercising on the target endpoint.
This test naming approach extends the previous convention by including the query parameter names on specific scenarios.
Arbitrary parameter values (e.g., the number ``42'' or the string ``fooBar'') are excluded, as such identifiers produce test names that are unnecessarily long and difficult to interpret.
Instead, this naming strategy uses only parameter names in one of the following scenarios
\begin{itemize}
\item the parameter value is the \texttt{true} boolean constant,
\item the parameter value is a negative numeric value (i.e., \texttt{-1},\texttt{-2}, etc.), or
\item the parameter value is an empty string value (i.e., ``'').
\end{itemize}
When the parameter name is included due to a negative numeric value, the token ``\texttt{negative}'' is also added to the parameter name as a prefix (e.g., \texttt{negativePrice}).
Similarly, when the parameter name is included because of an empty string parameter value, the token ``\texttt{empty}'' is additionally inserted as prefix.
For the test case example in Figure~\ref{fig:rest-assured-test} we have the following parameters and its values: \texttt{includeItems=true} and \texttt{currency=EUR}.
Since only \texttt{includeItems=true} satisfies one of the aforementioned scenarios (namely, the parameter value is the \texttt{true} boolean constant), the token ``\texttt{includeDetails}'' is added with no further prefixes.

This leads to the following test name for the test case in Figure~\ref{fig:rest-assured-test}:
\begin{center}
``\texttt{test\_0\_getOnUsersOrdersWithQueryParams\underline{IncludeDetails}ReturnsObject}''
\end{center}

Again, we underline the segments of the test case that are added with respect to the previously introduced \evoqueryshort test naming approach.

If we consider the \evo generated test case shown in Figure~\ref{fig:evomaster-generated-test}, we have that,
(i) the focal HTTP action is a \texttt{get} action,
(ii) the name qualifier in the path is \texttt{news}, with no parent qualifier,
(iii) the distinctive condition for the query parameters is that the value for parameter ``\texttt{country}'' is the empty string,
(iv) the expected result an empty list since the test case asserts that the JSON response body ``\texttt{size()}'' should be equal to $0$, and finally,
(v) the test case directly inserts data into the backend \texttt{SQL} database.
Therefore, the resulting test name with \evocondition test naming approach (abbreviated as \evoconditionshort) will be the following:
\begin{center}
``\texttt{test\_0\_getOnNewsWithQueryParamsEmptyCountryReturnsEmptyListUsingSQL}''
\end{center}

As previously stated, test cases do not exist in the vacuum, they are organized within test classes.
Additionally, another rule that the naming proposal adopt to reduce the length of the test case names is considering how frequent the \textit{\textless mechanism\textgreater} is present within the same test class.
If more than half of the test cases within the test class rely on the same stubbing/mocking/data insertion mechanism, we simply discard that \textit{\textless mechanism\textgreater} since it is redundant to understand the semantics of the test case in the context of the test class.
Finally, if the number of test class the final test name for \texttt{test\_3} from Figure~\ref{fig:evomaster-generated-test} under the previously described scenario will be:
\begin{center}
``\texttt{test\_0\_getOnNewsWithQueryParamsEmptyCountryReturnsEmptyList}''
\end{center}

\begin{figure}[!t]
\centering
\begin{lstlisting}
@Test
public void test2() {
    given().
        baseUri(baseUrlOfSut).
    when().
        get("/users/42/orders/1234?includeItems=true&currency=USD").
    then().
        statusCode(200).
        assertThat().
        contentType("application/json").
        body("currency", containsString("USD"));
}
\end{lstlisting}
\caption{\RESTAssured test checking currency in the response. The test case is similar to the one listed in Figure~\ref{fig:rest-assured-test} but the value of the currency query parameter is ``\texttt{USD}'' instead of ``\texttt{EUR}''. }
\label{fig:rest-assured-test-2}
\end{figure}


\subsection{LLM-based Approaches}
\label{sec:llm-based}

As discussed in Section~\ref{sec:mlapproaches}, Biagiola et al.~\cite{Biagiola_2025} showed how LLMs can be effectively prompted to improve test readability.
To generate test names with LLMs, we built upon the replication package provided by the authors~\cite{readability_llms}.
We adapted their readability-improvement tool so that it supplies the available OpenAPI schema to each LLM, and we instructed the models to \emph{only} produce descriptive names for REST APIs test cases.
The modified prompt we employed (with removed text as strikethrough and added text is underlined) is shown in Figure~\ref{fig:prompt}.

\begin{figure}[!t]
\begin{tcolorbox}
Improve the readability of the test below by modifying \ul{ONLY} \st{the identifiers}, the test name, \st{and variable name}, NOT THE FUNCTIONS CALLED INSIDE THE TESTS, STATIC METHOD, CALLED STATIC CLASS, \ul{VARIABLE NAMES NOR IDENTIFIERS}.\\

\ul{You must use both the API spec provided before, along with the test case code to determine a more descriptive test case name.}\\
\ul{When naming, you should not mention the goal of readability improvement, we're looking for real case test name improvement.}\\
\ul{Remember in test  naming it is important that:}\\
\ul{- Names be descriptive, there must be a relationship between the written code and the test case name.}\\
\ul{- Names should distinguish uniquely a test case from another.}\\
\ul{- Names allow us to understand which part of the source code is being tested. In this case, which part of the API described by the API spec.}\\
The changes must not affect the functioning of the test in any way.
\end{tcolorbox}
\caption{\label{fig:prompt}
LLM prompt, modified from~\cite{readability_llms}.
Removed text is marked with strikethrough, whereas added text is underlined.
}
\end{figure}

\subsection{Test Ordering}
\label{sec:test_ordering}

\evo generates independent test cases, meaning that their execution order does not affect their outcomes.
As a result, the generated test cases can be freely reordered, grouped into different test suites or files, and executed either in isolation or toghether with other test cases~\cite{marculescu2022faults}.

 By default, \evo sorts test cases by prioritizing those whose HTTP calls return error codes (i.e., status codes in the \texttt{5xx} range), followed by those returning \texttt{2xx} status codes, and finally those returning \texttt{4xx}.
 Within each status-code group, tests are further ordered by the number of covered targets: the more targets a test case covers, the higher it appears in the ordering.

In this article, we introduce a novel ordering for \evo generated test cases.
This ordering follows the intuition that, when testing a API with multiple paths such as \texttt{/users} and \texttt{/users/\{userId\}/permissions}, it makes sense to first assure that the behaviour of \texttt{/users} is correct before testing more specific endpoint such as the user's permission, regardless of how many targets each test covers.

The new test case ordering follows three levels of priority:
\begin{enumerate}
\item \textbf{Paths:} tests are grouped and ordered by their path, prioritizing more general endpoints over more spefic ones.
\item \textbf{HTTP status codes:} within each path group, we sort test cases as previously described.
Those whose HTTP calls return status codes in the \texttt{5xx} range appear first, followed by those returning \texttt{2xx} status codes, and finally those returning a \texttt{4xx} status code.
\item \textbf{HTTP verbs:} finally, test cases are ordered by the HTTP method in the following order: \texttt{GET}, \texttt{POST}, \texttt{PUT}, \texttt{DELETE}, \texttt{OPTIONS}, \texttt{PATCH}, \texttt{TRACE} and \texttt{HEAD}.
This order reflects a logical, human-like testing progression—similar to what one would do in Test-Driven Development (TDD)~\cite{santos2021family}: first checking retrieval (\texttt{GET}), then creation (\texttt{POST}), modification (\texttt{PUT}), deletion (\texttt{DELETE}), and so on.
\end{enumerate}

Note that when two test cases have identical \textbf{path}, \textbf{HTTP status code}, and \textbf{HTTP verb}, their relative order remains unchanged, reflecting the order in which they were originally created.
We propose these criteria based on practical development experience, aiming to make the ordering more intuitive and human-like, reflecting how a developer would naturally structure tests.

\section{Evaluation}
\label{sec:evaluation}

To evaluate the effectiveness of the proposed techniques for generating descriptive names of automatically generated \RESTAssured system tests, we conducted an empirical evaluation aimed at answering the following research questions:

\begin{itemize}
\item RQ\#1: Which of the considered approaches to test naming is best suited for test cases generated by \evo?
\item RQ\#2: What conclusions do developers take of the best choice for generating test names within \evo?
\end{itemize}

\subsection{Experimental Setup}

\subsubsection{Surveys}

In order to answer RQ\#1, we prepared two surveys for our participants to complete.
We refer to each survey as Survey \#1 and Survey \#2.
Surveys were conducted via Google Forms~\cite{googleForms}.
Each survey presented five different test cases, each displayed as an image containing its source code, along with a question asking for the name of the test case.
We used two distinct surveys of five test cases each, instead of a single survey with 10 test cases, to avoid fatigue in the respondents and make them more likely to accept to participate in this study.
If the respondents had time and interest, they were asked to complete both surveys.

\begin{figure}[!t]
\begin{center}
        \includegraphics[width=0.5\linewidth]{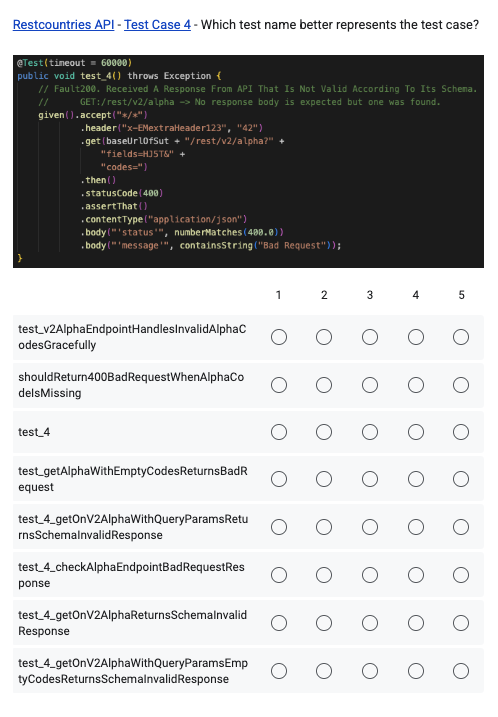}
           \caption{Sample test case question for REST API \emph{RestCountries} listed in Survey \#1.}
           \label{fig:survey-question}
\end{center}
\end{figure}

Figure~\ref{fig:survey-question} presents one of the displayed tests cases in Survey \#1.
If the image was not large enough for the participant to read comfortably, they could find the test source code in a Java file in a \GitHub repository.
For each test case, we asked each participant to rate a list of presented test names on a scale from $1$ to $5$, with $1$ being the least representative or helpful and $5$ being the most.
To mitigate potential bias due to respondent fatigue, the order of test cases and the test names was randomized for each participant within the same survey.
We encouraged participants to submit their answers, even if the survey was not fully completed due to fatigue.

\subsubsection{Subjects}

\begin{figure}[!t]
    \centering
    \begin{subfigure}{0.5\textwidth}
        \includegraphics[width=\linewidth]{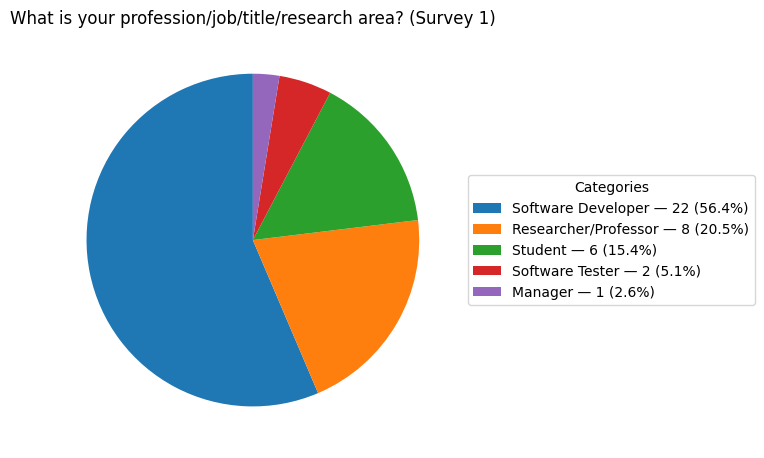}
           \caption{Professional roles of the $39$ answers in Survey \#1}
    \end{subfigure}%
    \begin{subfigure}{0.5\textwidth}
        \includegraphics[width=\linewidth]{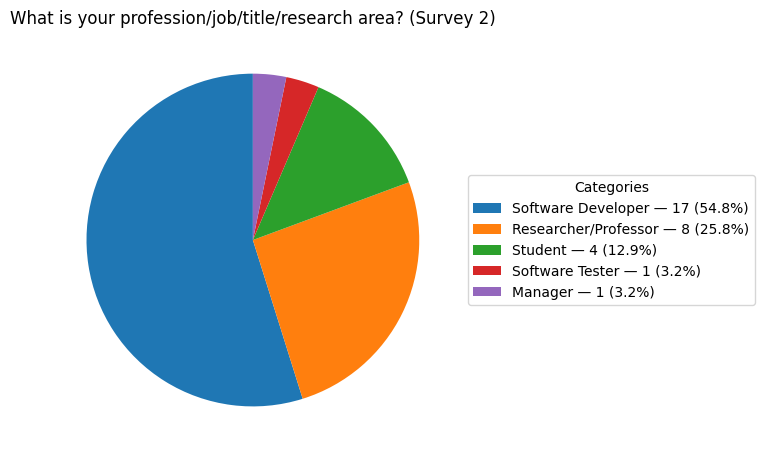}
           \caption{Professional roles of the $31$ answers in Survey \#2}
    \end{subfigure}
    \caption{Answers to \emph{What is your profession/job/title/research area?} question in surveys \#1 and \#2}
    \label{fig:professional_roles}
\end{figure}

\begin{figure}[!t]
    \centering
    \begin{subfigure}{0.5\textwidth}
        \includegraphics[width=\linewidth]{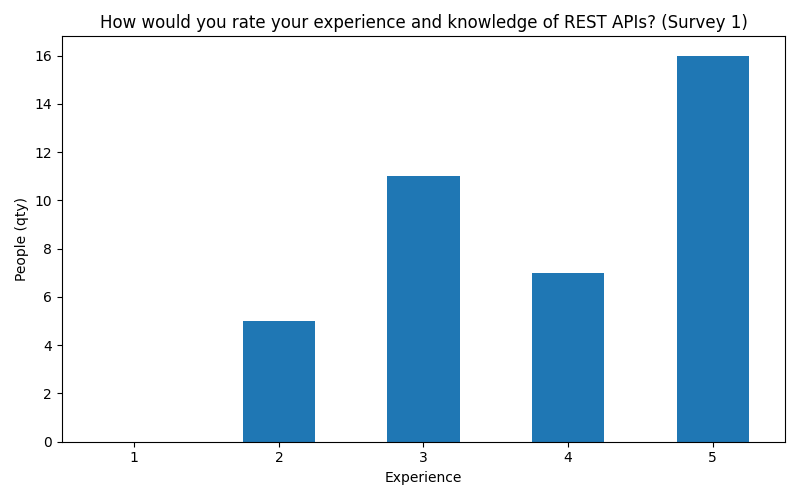}
           \caption{Survey \#1 REST API knowledge}
    \end{subfigure}%
    \begin{subfigure}{0.5\textwidth}
        \includegraphics[width=\linewidth]{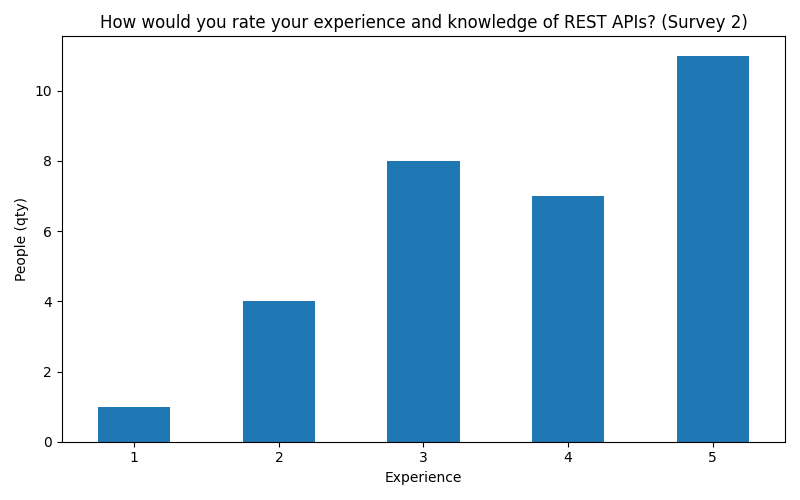}
           \caption{Survey \#2 REST API knowledge}
    \end{subfigure}
    \caption{Answers to \emph{How would you rate your experience and knowledge of REST APIs?} question in surveys \#1 and \#2.
    The scale goes from 1=Not Experienced to 5=HighlyExperienced.}
    \label{fig:rest_api_knowledge}
\end{figure}

\begin{figure}[!t]
    \centering
    \begin{subfigure}{0.5\textwidth}
        \includegraphics[width=\linewidth]{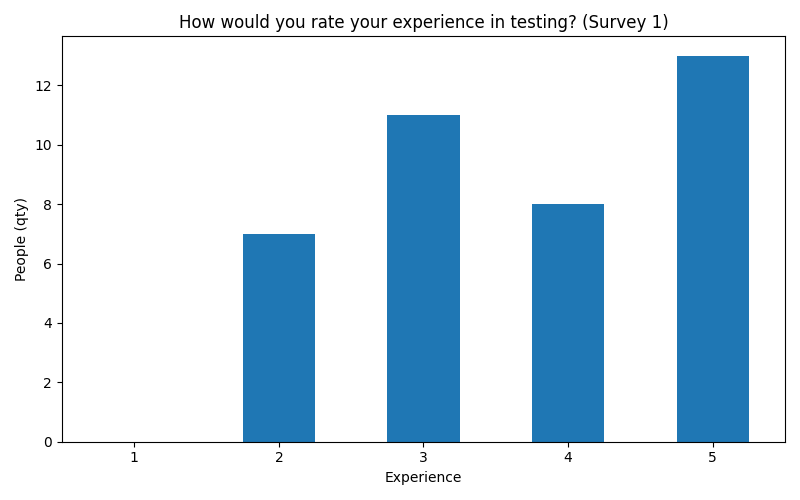}
           \caption{Survey \#1 testing knowledge}
    \end{subfigure}%
    \begin{subfigure}{0.5\textwidth}
        \includegraphics[width=\linewidth]{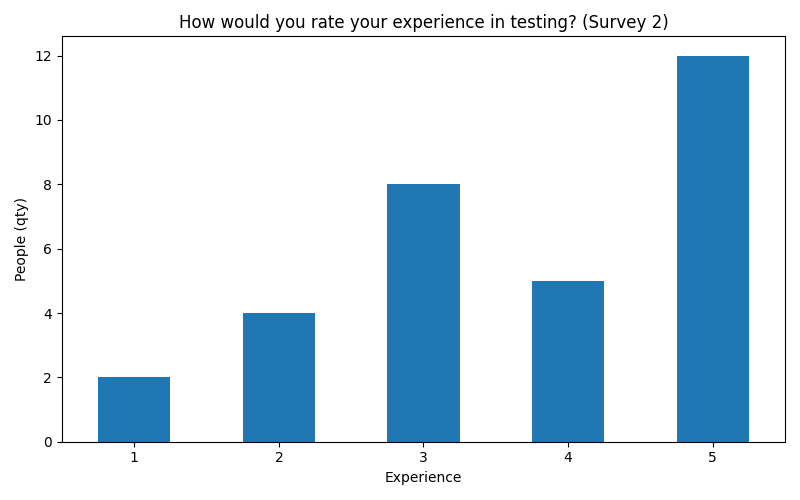}
           \caption{Survey \#2 testing knowledge}
    \end{subfigure}
    \caption{Answers to \emph{How would you rate your experience testing?} question in surveys \#1 and \#2.
    The scale goes from 1=Not Experienced to 5=HighlyExperienced.}
    \label{fig:testing_knowledge}
\end{figure}

We recruited participants by publishing a call on \LinkedIn, targeting individuals with software-related backgrounds,  including both researchers and industry professionals.
The announcement specified the purpose of the survey and included an estimated time commitment ($10$ minutes).
To characterize our participants, we asked them to report their professional role (Figure~\ref{fig:professional_roles}).
We also asked their perceived expertise in REST APIs (Figure~\ref{fig:rest_api_knowledge}) and their perceived testing skills (Figure~\ref{fig:testing_knowledge}).
The answer to both questions was provided on a 5-point Likert scale~\cite{likert1932technique} (being 1 = Not experienced, 5 = Highly experienced).

In total, we received $39$ answers to Survey \#1, while $31$ answers were received for Survey \#2.
In other words, $31$ people had time and willingness to complete both surveys, whereas $8$ only completed the first survey.
Most of these participants were software developers stating having good knowledge of REST APIs and testing.

\subsubsection{Test Case Selection for Surveys}

We followed a systematic protocol to create a sample of $10$ generated test cases for our study, using a subset of JVM-based REST APIs from the WebFuzzing Dataset~\cite{sahin_2025_wfc}\textemdash a curated collection of Web APIs commonly used in software testing research.
For practical reasons, we restricted our selection to the $13$ JDK 8/11 REST APIs built with Maven~\cite{Maven}.
Our selection protocol consisted of the following steps:
\begin{enumerate}
\item \textbf{Generation}: We generated test cases using \evo in white-box mode with a budget of $15$ minutes for each of the $13$ chosen REST APIs.
\item \textbf{Filtering}: We selected the generated test cases where its last HTTP call contained any query parameter whose value was the \texttt{true} boolean constant, a negative number or the empty string.
This yielded $71$ test cases across $9$ different REST APIs.
\item \textbf{Sampling}: We performed stratified random sampling by selecting one test case uniformly at random from each of the $9$ REST API-specific subsets.
Since the number of REST APIs was less than our target of sample size of $10$ test cases, we completed the sample by randomly choosing a $10^{th}$ generated test case among the remaining $71-9=62$ filtered test cases.
\end{enumerate}

\subsubsection{Treatments}

We selected a range of treatments for our survey focusing on \RESTAssured test name generation.
We first included \numbered which is the default test naming approach currently used in \evo.
This approach simply generates test names as ``\texttt{test0}'', ``\texttt{test1}'', etc.
Secondly, we considered the three rule-based approaches introduced in Section~\ref{sec:rule-based}: \evosimpleshort, \evoqueryshort and \evoconditionshort.

As representatives of the LLM-based approaches, we considered those reported as the most effective LLMs by Biagiola et al.~\cite{Biagiola_2025}: \emph{Google DeepMind Gemini}~\cite{gemini_2025} (version 1.5-pro-001), and \emph{OpenAI GPT}~\cite{Brown_2020} (versions 3.5 and 4o) with the adapted prototype described in Section~\ref{sec:llm-based}.

To study how providing the whole REST API source code, rather than only the OpenAPI schema, affects the test naming generation, we additionally considered \GitHub Copilot~\cite{ziegler2024measuring}.
 The procedure for generating test names with \GitHub Copilot was the following:
we first opened each REST API project with Microsoft Visual Studio Code~\cite{VsCode}, an integrated development environment (IDE) with support for \GitHub Copilot.
Secondly, we select the targeted test case, prompting an adapted query as the one given to the other LLMs.
\GitHub Copilot relies on GPT-4o as the backend LLM.

\subsection{Survey Results}

\begin{figure}[!t]
    \centering
        \begin{subfigure}{0.5\textwidth}
        \includegraphics[width=\linewidth]{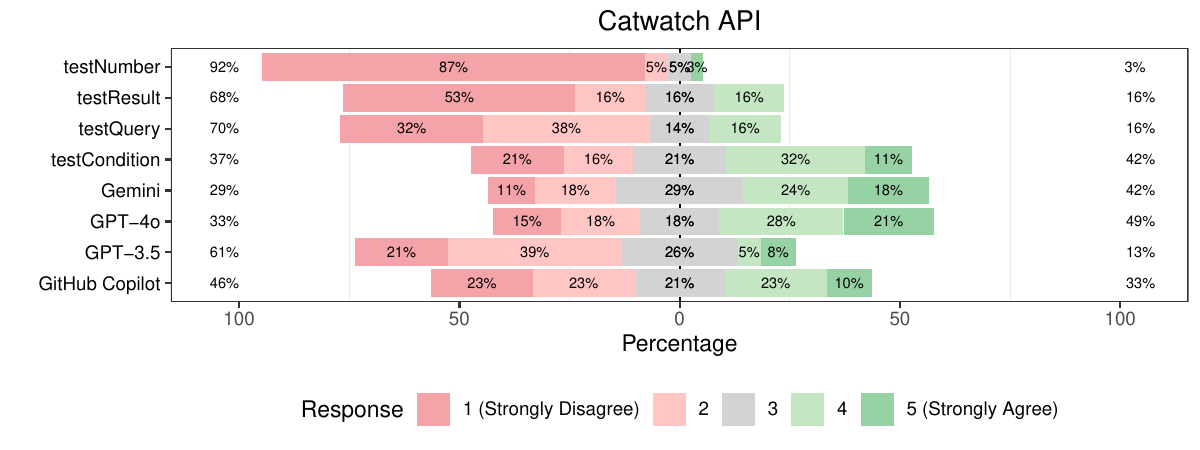}
           \caption{catwatch}
    \end{subfigure}%
        \begin{subfigure}{0.5\textwidth}
        \includegraphics[width=\linewidth]{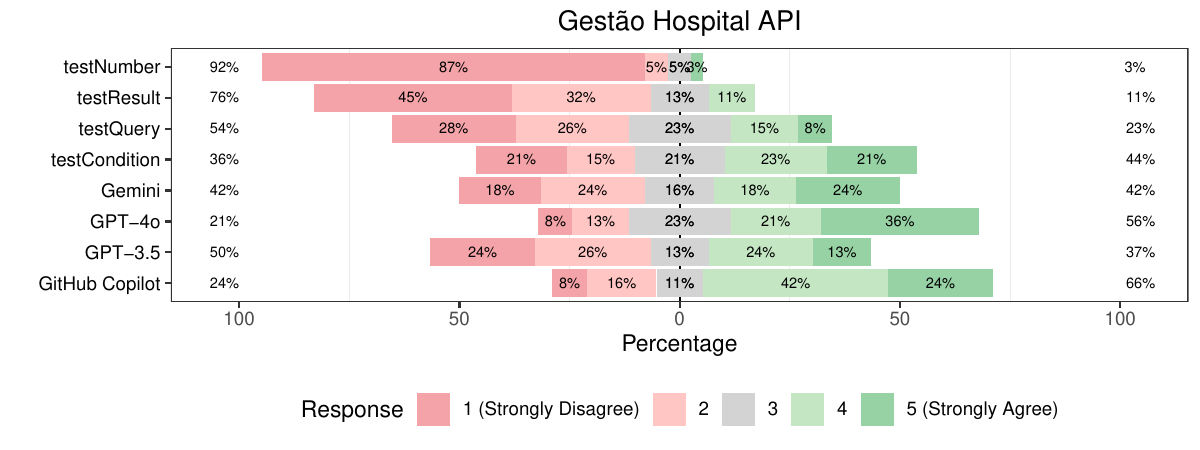}
        \caption{gestaohospital}
    \end{subfigure}
	\\
    \begin{subfigure}{0.5\textwidth}
        \includegraphics[width=\linewidth]{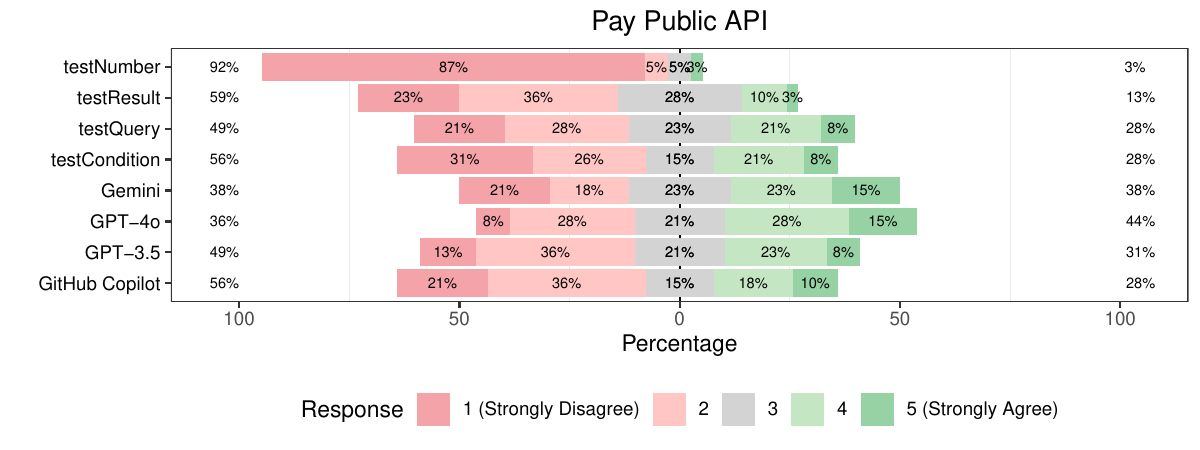}
        \caption{pay-publicapi}
    \end{subfigure}%
    \begin{subfigure}{0.5\textwidth}
       \includegraphics[width=\linewidth]{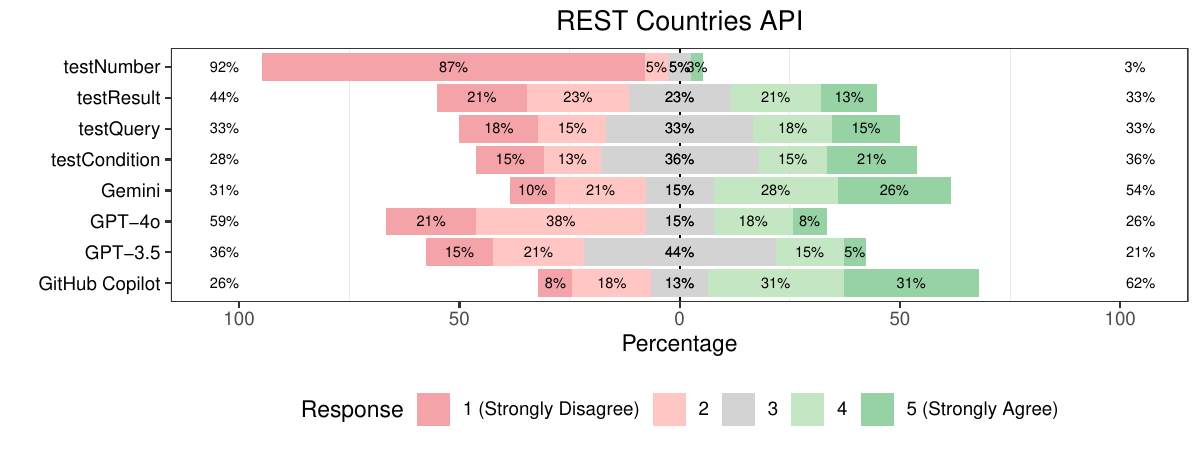}
        \caption{restcountries}
    \end{subfigure}
    \\
    \begin{subfigure}{0.5\textwidth}
        \includegraphics[width=\linewidth]{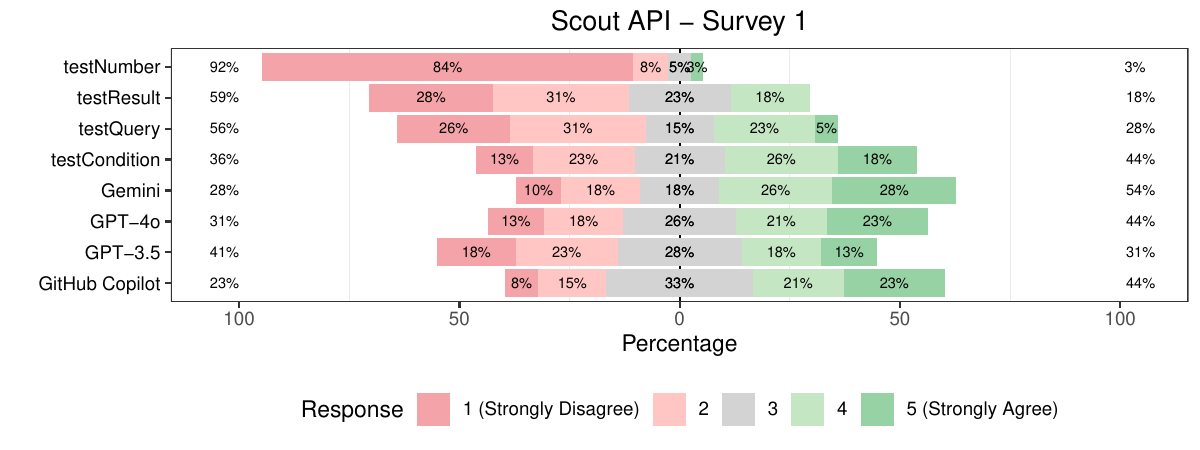}
        \caption{scout-api Survey \#1}
    \end{subfigure}
    \caption{Likert-scale results for selected test cases in Survey \#1.}
    \label{fig:likert_plot_survey_1}
\end{figure}

\begin{figure}[!ht]
    \centering
    \begin{subfigure}{0.5\textwidth}
        \includegraphics[width=\linewidth]{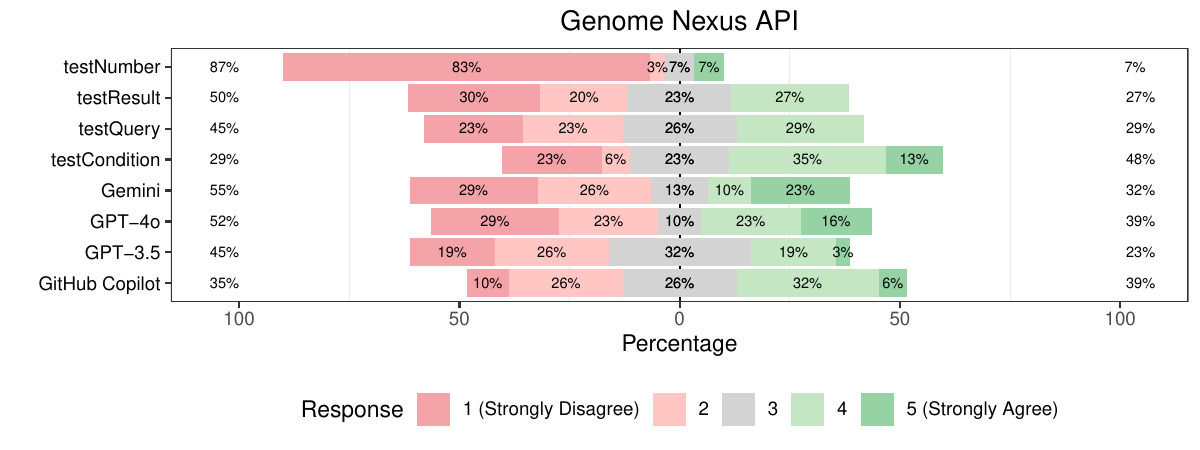}
        \caption{genome-nexus}
    \end{subfigure}%
    \begin{subfigure}{0.5\textwidth}
        \includegraphics[width=\linewidth]{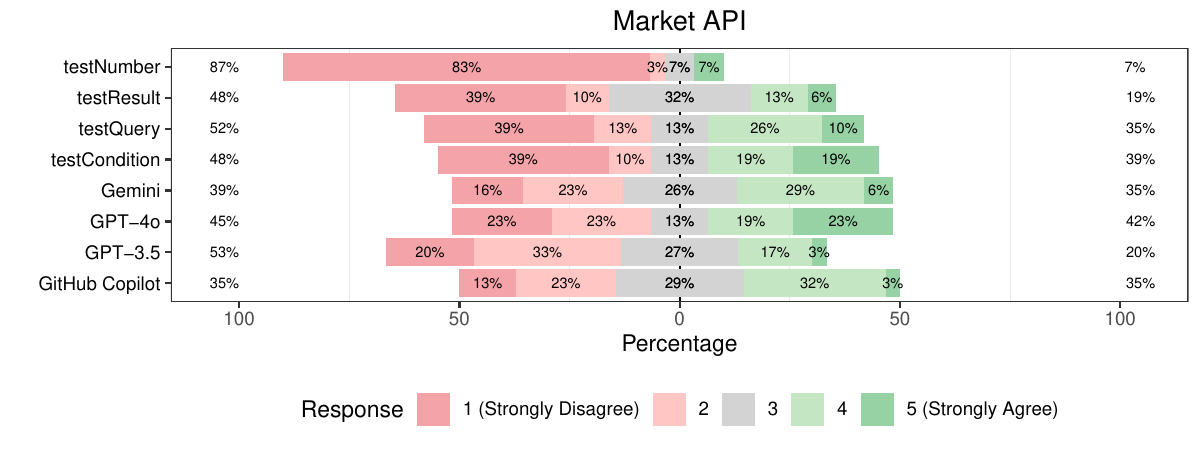}
        \caption{market}
    \end{subfigure}
    \\
    \begin{subfigure}{0.5\textwidth}
        \includegraphics[width=\linewidth]{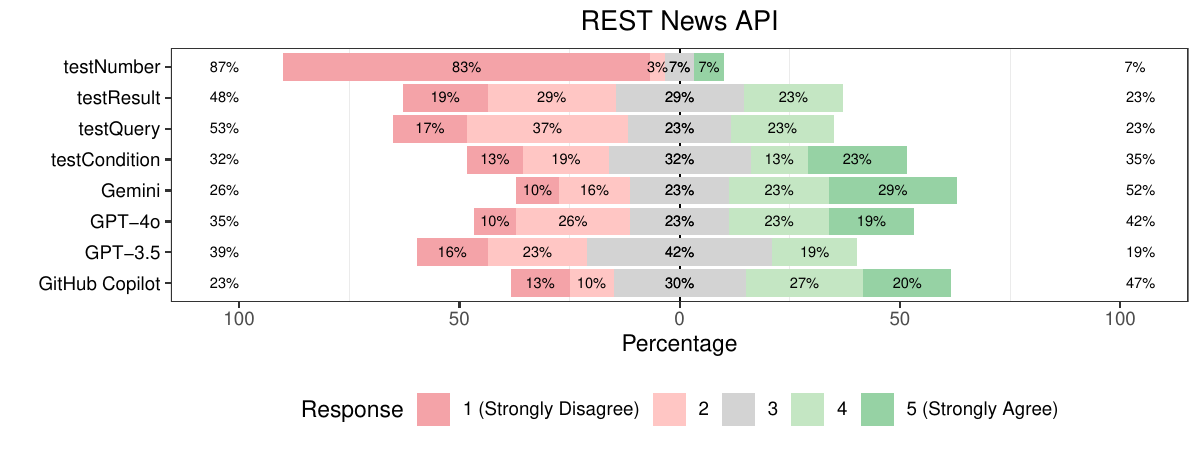}
        \caption{rest-news}
    \end{subfigure}%
    \begin{subfigure}{0.5\textwidth}
        \includegraphics[width=\linewidth]{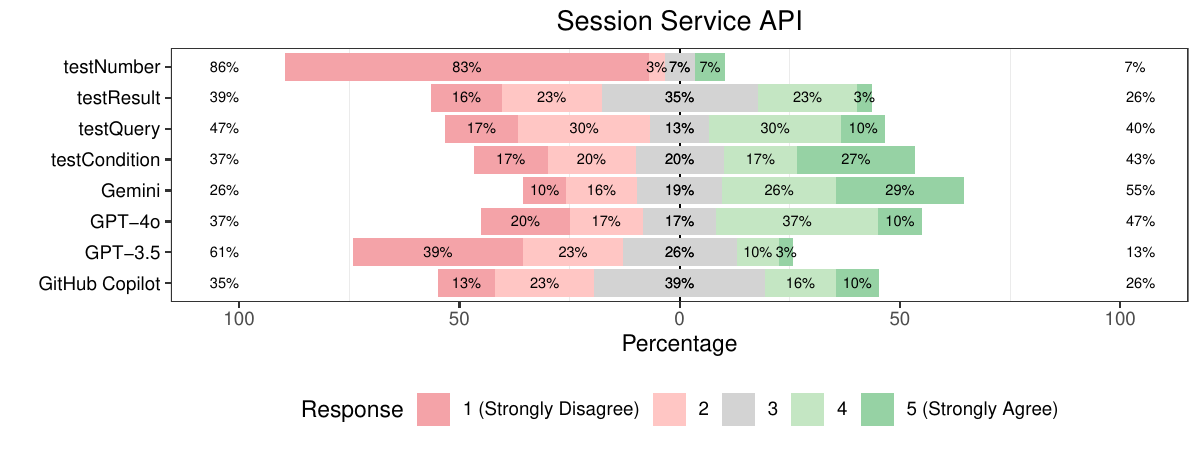}
        \caption{session-service}
    \end{subfigure}
    \\
    \begin{subfigure}{0.5\textwidth}
        \includegraphics[width=\linewidth]{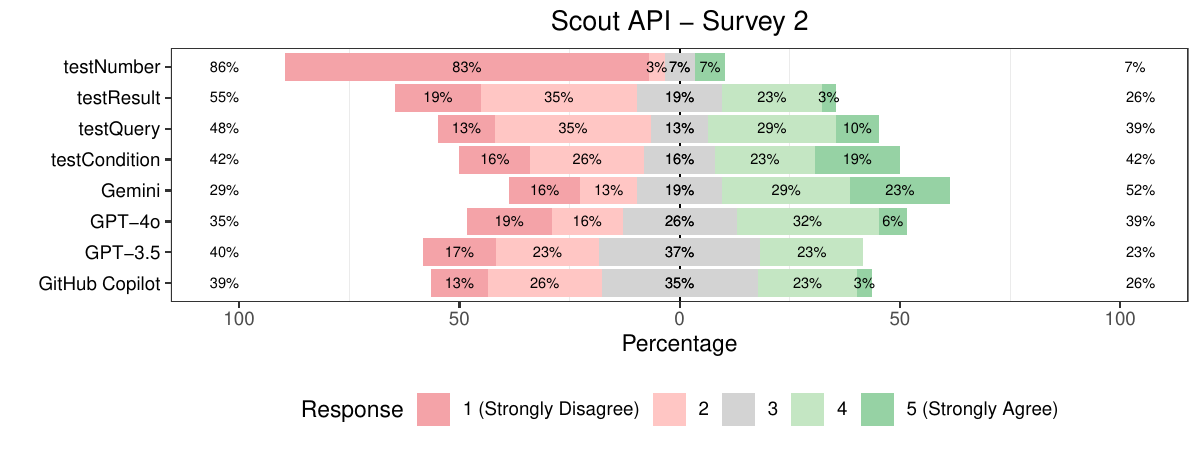}
        \caption{scout-api  Survey \#2}
    \end{subfigure}
    \caption{Likert-scale results for selected test cases in Survey \#2.}
    \label{fig:likert_plot_survey_2}
\end{figure}

Figures~\ref{fig:likert_plot_survey_1} and~\ref{fig:likert_plot_survey_2} show the answers given by the participants on each survey.
Table~\ref{tab:statistical_significance} presents the comparison of each test generation treatment, pair comparisons per SUT are analyzed with Wilcoxon-Mann-Whitney U-tests, with $\hat{A}_{12}$ Vargha-Delaney standarized effect-sizes.
To verify the difference in performance among the tools, a Friedman Test is carried out.

\begin{figure}[!t]
\begin{center}
\includegraphics[width=13cm]{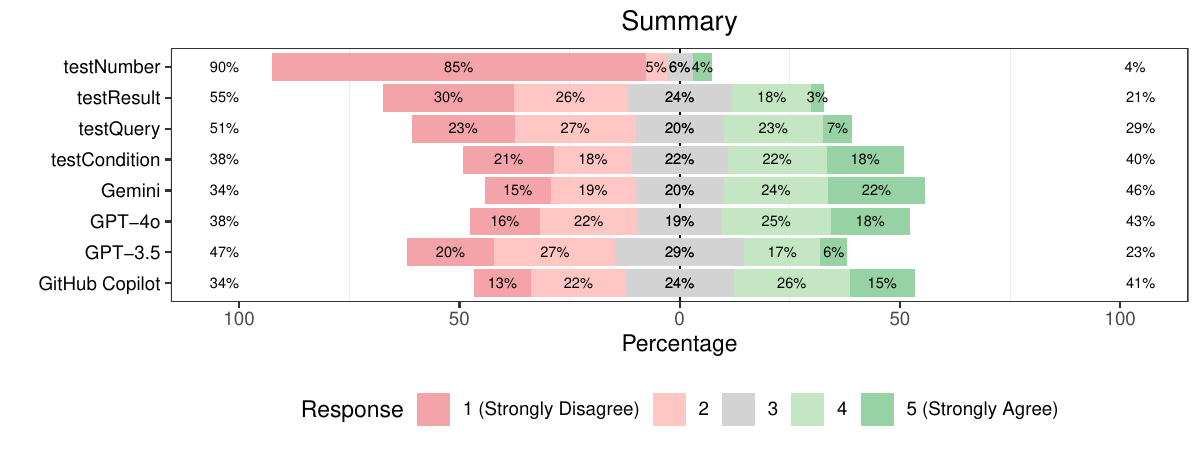}
\end{center}
    \caption{All plots combined into one figure.}
    \label{fig:overall}
\end{figure}

We start by studying the ratings for the default \evo test naming approach (i.e., \numbered) and the three proposed rule-based test naming approaches —\evosimpleshort, \evoqueryshort and \evoconditionshort.
Among these four treatments, participants consistently preferred the \evoconditionshort approach in both surveys.
In Survey \#1, positive responses ranged from $28\%$ (for subject \texttt{pay-publicapi}) to $44\%$ (for subjects \texttt{scout-api} and \texttt{gestaohospital}).
Notably, only for a single subject the number of participants expressing positive feedback matched between \evoconditionshort and \evoqueryshort.
In Survey \#2, across all five subjects, participants consistently rated \evoconditionshort more positively than the other rule-based approaches.
For example, in subject \texttt{genome-nexus}, $48\%$ of participants gave positive feedback for the name generated by \evoconditionshort, compared to $29\%$ for \evoqueryshort and $27\%$ for \evosimpleshort.
When studying the effect sizes presented in Table~\ref{tab:statistical_significance}, we observe that \evoconditionshort achieves higher effect sizes ($0.43$ and $0.38$) with statistical significance.
It is worth noticing that \evoconditionshort outputs test names that are larger than both \evoqueryshort and \evosimpleshort.
This means that most participants found longer test names more declarative than shorter ones.
Figure~\ref{fig:overall} shows the overall reviews achieved by all treatments among all subjects for both surveys.
Overall, \evoconditionshort got $40\%$  positive reviews, while \evoqueryshort and \evosimpleshort achieved $29\%$ and $21\%$ respectively.

\begin{result}
{\bf RQ\#1}:  The \evoconditionshort test naming approach was the most positive rated among the different rule-based approaches.
The average effect-size is $\hat{A}_{12}=0.40$, and overall it achieved a $40\%$ of positive reviews.
\end{result}

Results are more mixed when comparing the best performing rule-based test naming approach (i.e., \evoconditionshort) with the LLM-based test name generation.
We observe from Figure~\ref{fig:likert_plot_survey_1} that for Survey \#1 \evoconditionshort does not achieve the highest ratings (values 4 and 5 on the Likert scale) in any subject.
When considering the ratings for Survey \#2 shown in Figure~\ref{fig:likert_plot_survey_2}, we observe that \evoconditionshort  achieves the highest ratings for only one of the subjects (i.e., \texttt{genome-nexus}).
Despite these descriptive trends, the pairwise comparisons reveal no statistically significant effect-size differences between \evoconditionshort, \emph{Gemini}, \emph{GPT-4o}, or \emph{\GitHub Copilot}.
In contrast, pairwise comparisons indicate that \emph{GPT-3.5} receives statistically significantly lower ratings compared to \evoconditionshort and all the other LLM-based approaches .

\begin{result}
{\bf RQ\#1}: The differences in ratings between \evoconditionshort and all LLM-based approaches, with the exception of \emph{GPT-3.5}, are not statistically significant.
However, \evoconditionshort achieves statistically significant higher ratings than \emph{GPT-3.5}.
\end{result}

After establishing that the empirical study revealed no statistically significant differences among \evoconditionshort, \emph{Gemini}, \emph{GTP-4o} and \emph{\GitHub Copilot}, we concluded that \evoconditionshort is the most suitable choice for the \evo generated test cases.
We followed the rationale that, when choosing between a deterministic, lightweight rule-based approach such as \evoconditionshort and LLM-based approaches—which are environmentally costly~\cite{Niu_2025} and may raise security and privacy concerns~\cite{Chandra_2025}—the former is preferable.

Exact comparisons with LLM approaches are hard to carry out in a fair and precise manner~\cite{sartaj2025searchbasedsoftwareengineeringai}.
A deterministic approach like \evoconditionshort can run in the same process of the fuzzer, directly using the internal data structures of the test cases to quickly compute a name.
That takes fractions of milliseconds.
On the other hand, using an LLM would require making a call over the network, compute the inference on a high-end GPU, and send back the response over the network.
That can easily take over a second.
If this is repeated for each generated test case, in a test suite with hundreds of tests this overhead can become in the order of minutes.
This can have detrimental effects on the search (e.g., given the same time budget, less test cases can be evaluated during the fuzzing process).
Furthermore, high-end GPUs consume significantly more energy than CPU cores.
While a CPU handles \evoconditionshort in mere milliseconds, an LLM can take several seconds for inferencing.
A server CPU core typically uses around 4W of power~\cite{poth2025eco}.
In contrast, an NVidia A100 GPU under load consumes approximately 300W, and deploying larger LLMs would require several A100 GPUs. Although we cannot provide precise measurements, it is evident that LLM approaches are at least two orders of magnitude slower and more energy-intensive than \evoconditionshort.

Besides speed and cost, there are two further issues with the use of LLMs in this context.
With LLMs, there is no guarantee that no \emph{hallucination} will take place.
Even if there is no hallucination, test engineers would still need to look at and validate each generated test name to make sure that's the case.
This takes time.
On the other hand, a deterministic approach is more reliable in its results, and prevents hallucinations.

A second major issue is \emph{consistency}.
If a fuzzer generates as output 200 test cases (for example), even by using the same LLM prompt there is no guarantee that the different generated names will follow the same ``style''.
Differences in styles might increase the cognitive load for the test engineers, making the generated test suite harder to read and understand.
This problem would not happen with a deterministic approach.
Although our experiment design does not enable us to study and quantify the effects of this potential issue, this is a further argument in favour of using deterministic approaches, unless LLMs can provide  better results.
But this was not the case in our experiments.

One potential benefit of LLMs though is that, by taking into account the whole OpenAPI schema, and the description comments in it (if any), LLMs might be able to include semantic information when generating test names.
This could lead to better, more meaningful test names.
We have not seen such occurrence in our experiments, but that does not mean it could not happen with different prompts or with new LLMs in the future.

\begin{result}
{\bf RQ\#1}: \evoconditionshort is the preferred test naming approach for \evo-generated test cases, ranking on par with LLMs while offering lower environmental and security risks.
\end{result}

\begin{table}[!t]
    \caption{The Vargha-Delaney $A_{12}$ measure for each treatment across the 10 test cases.
    Comparisons where the Mann-Whitney-Wilcoxon U-tests ($p$-$value$) are lower than $<0.05$ are presented in bold.
    }
    \label{tab:statistical_significance}
\begin{center}
\begin{tabular}{l|cccccccc}
 &  \makecell{test \\ Number} & \makecell{test \\ Result} & \makecell{test \\ Query} & \makecell{test \\ Condition} & Gemini & GPT-4o & GPT-3.5 & \makecell{GitHub \\ Copilot}  \\
\hline
testNumber & 0.50 & {\bf 0.25} & {\bf 0.22} & {\bf 0.20} & {\bf 0.17} & {\bf 0.18} & {\bf 0.21} & {\bf 0.16} \\
testResult & {\bf 0.75} & 0.50 & {\bf 0.45} & {\bf 0.38} & {\bf 0.34} & {\bf 0.36} & {\bf 0.44} & {\bf 0.35} \\
testQuery & {\bf 0.78} & {\bf 0.55} & 0.50 & {\bf 0.43} & {\bf 0.39} & {\bf 0.41} & 0.50 & {\bf 0.40} \\
testCondition & {\bf 0.80} & {\bf 0.62} & {\bf 0.57} & 0.50 & 0.46 & 0.49 & {\bf 0.57} & 0.48 \\
Gemini & {\bf 0.83} & {\bf 0.66} & {\bf 0.61} & 0.54 & 0.50 & 0.53 & {\bf 0.61} & 0.52 \\
GPT-4o & {\bf 0.82} & {\bf 0.64} & {\bf 0.59} & 0.51 & 0.47 & 0.50 & {\bf 0.59} & 0.50 \\
GPT-3.5 & {\bf 0.79} & {\bf 0.56} & 0.50 & {\bf 0.43} & {\bf 0.38} & {\bf 0.41} & 0.50 & {\bf 0.40} \\
GitHub Copilot & {\bf 0.84} & {\bf 0.65} & {\bf 0.60} & 0.52 & 0.48 & 0.50 & {\bf 0.60} & 0.50 \\
\hline
Friedman Test & \multicolumn{8}{r}{\textbf{$\chi^2$ = 535.234, $p$-value = $\le $0.001} } \\ 
\end{tabular}

\end{center}
\end{table}

\subsection{Industry Study Results}

In the surveys \#1 and \#2, participants were not familiar with the REST APIs under test and may not have been familiar with \evo.
In order to answer research question RQ\#2 on how practitioners (e.g., developers and testers) react to the proposed rule-based testing approach for test cases generated by \evo,
we aimed at practitioners that were both familiar with \evo and the REST APIs under test.
We prepared a brief questionnaire of seven questions on how does this \evoconditionshort test naming approach compares to the previous default of \evo (i.e., \numbered).

As \evo is actively used at Volkswagen AG~\cite{poth2025technology,icst2025vw}, we asked the test engineers at Volkswagen AG currently generating test cases on their REST APIs using \evo to complete a brief questionnaire and provide their insights on the overall test naming approach.
We asked these participants to execute \evo again with its latest (at the time of writing) version $4.0.0$, which has the \evoconditionshort approach as the default, and compare the results with their experience using the \numbered test naming technique used in the earlier versions of \evo.
They applied \evo on two distinct APIs, and analyzed the results for every single generated test case.
We received four answers to the questionnaire from test engineers at Volkswagen AG.

Note that, besides providing a new way to name test cases, we also provided a novel approach to sort test cases within the same generated test suite presented in Subsection~\ref{sec:test_ordering}.
One question (Q6) in the questionnaire was specific about this new test order mechanism.
Notice that this ordering was not evaluated in the previous empirical study as the test cases were introduced in isolation (i.e., one test case for each subject).

Of the seven questions, three were closed-ended answers while the remaining four were open-ended answers.
We start by presenting the answers to the first three closed-ended answers:

\begin{itemize}
\item[(Q1)] \emph{Do you prefer this new naming convention (now the default in \evo) or the old one based on numbering (e.g., \texttt{test\_01()})?}\\
All four respondents answered that the \evoconditionshort test naming approach was preferable to the previous \numbered test naming technique.\\

\item[(Q2)] \emph{On a scale from 1 to 5, where 5 is best, how do you rank the old naming convention and the new convention?
   Rank them separately, i.e., old=[1-5] and new=[1-5].}\\
 In the 1 to 5 scale, all four participants ranked \evoconditionshort with a score of "4" (Good), while the previous \numbered test naming received a score of  "1" (Worst) or "2" (Poor).\\

\item[(Q3)] \emph{Out of $N$ generated test cases, how many:
   ``g'' have good enough names?
   ``c'' have names that require some minor changes?
   ``m'' have very confusing or misleading names?
   where $N$ = g + c + m.
   Provide values separately for each different tested API.
}\\
Across the four respondents, the quality of the generated test names was generally rated positively.
Respondent \#1 created $21$ test cases and considered all of the test names produced with \evoconditionshort to be \emph{good enough}.
Respondent \#2 generated $10$ test cases: $5$ names were judged \emph{good enough}, $3$ required some minor changes, and $2$ were deemed confusing or misleading. \\
Respondent \#3 produced $13$ test cases, with $9$ rated as \emph{good enough} and $4$ needing minor adjustments.
Finally, respondent \#4 (who generated the largest number of generated test cases, i.e., 30), found $29$ out of them \emph{good enough}, while only $1$  required minor modification.
Overall, the majority of generated test names were perceived as good enough, with only a few requiring minor refinement and very few being considered unclear.

\end{itemize}

The remaining four closed-ended questions in the questionnaire were the following:
\begin{itemize}
\item[(Q4)] \emph{Is there anything you think is missing or should be changed in these generated names?}

Most respondents considered the generated test names sufficiently informative and generally adequate.
However, several respondents identified missing semantic cues that could enhance interpretability.
In particular, they noted the absence of information regarding whether a test case represented a positive or negative scenario (respondent \#1), the lack of explicit HTTP status codes or expected outcomes (respondent \#2 and \#3), and occasional truncation of words within names (respondent \#4).

Overall, respondents suggested that the current naming structure could be improved through the inclusion of richer semantic details rather than through structural changes to \evoconditionshort itself. \\

\item[(Q5)] \emph{Can you give us some examples of a generated name for a test and how you would have rather named
   that test? If so, can you explain what concepts were missing in the generated name? (i.e., why
   would you need to change them)}

When asked to propose alternative names, respondents agreed on the idea of increasing semantic expressiveness.

Respondents \#1 and \#2 highlighted that test names ending with ``\texttt{Returns\-Schema\-Invalid\-Response}'' could be improved to make them more declarative.
In particular, these test suffixes are generated expanding the \textit{\textless faultLabel\textgreater} placeholder whenever the test case \emph{``received a response from API with a structure/data that is not matching its schema''}.
Such fault labels are listed in the Web Fuzzing Commons library~\cite{sahin_2025_wfc}.
Respondent \#3 similarly recommended including both the HTTP code and parameter context, transforming names such as \texttt{test\_0\_getOnSupplierReturnsObject()} into \texttt{test\_0\_get\-Supplier\-Returns\-200\-With\-All\-Mandatory\-Parameters()}.
Finally, respondent \#4 highlighted the need to avoid truncated tokens in identifiers (e.g., \texttt{postOnConsum} instead of \texttt{postOnConsume}).
This happened due to the length limit $N=80$.
After this feedback, we have modified its default value to $N=120$.

Collectively, these suggestions emphasize that respondents valued semantic precision, completeness, and readability over brevity.\\

\item[(Q6)] \emph{Do you have any comments on how the test cases are ordered/sorted inside the generated test suite files?
   (This is NOT related to test execution order, BUT rather on text position inside the Java files)}

All respondents reported that the current textual ordering of tests within generated files was acceptable.
One participant (respondent \#2) proposed enriching the generated test cases with semantic information
by adding \texttt{@Tag} annotations from OpenAPI\footnote{\url{https://github.com/OAI/OpenAPI-Specification/blob/main/versions/3.0.0.md\#tag-object}} to specify some desired test feature (e.g., the test case returns a specific type of object), which could be particularly useful for large test suites.
The remaining respondants indicated that ordering did not present any practical issues in their evaluations.\\

\item[(Q7)] \emph{Do you have any final high-level comments regarding the generated names and the test position sorting?}

Three participants (respondents \#1, \#2, and \#4) provided no further remarks.
Respondent \#3 reiterated the value of recognizable and descriptive names, as already discussed in their answer to Q5, to further improve clarity for developers and domain experts.

\end{itemize}

\begin{result}
{\bf RQ\#2}: Test engineers at Volkswagen AG concluded that \evoconditionshort is clearly the better choice for test names generated by \evo compared to the previous default \numbered  technique.
Some feedback highlighted that certain names could be made more declarative and enriched with semantic annotations.
\end{result}

\section{Threats To Validity}
\label{sec:threats_to_validity}

As with any empirical study, our findings are subject to several threats to validity.

\textbf{Internal validity.}
Threats to internal validity include learning and fatigue effects, as participants rated multiple test cases, as well as differences in how test cases were presented (images vs. source code files).
To mitigate these threats, we randomized the order of test cases and proposed names for each participant and provided access to the full Java files when images were too small.
We also collected participants' self-reported expertise in REST APIs and testing to contextualize potential individual differences.

\textbf{External validity.}
Threats to external validity concern the generalizability of the participants, the surveyed test cases, and the evaluation context.
To reduce participant-related bias, we recruited respondents through LinkedIn rather than selecting them arbitrarily.
Similarly, generated test cases were chosen using strict systematic protocol, including filtering and stratified random sampling, to avoid choosing them arbitrarily.
Because most survey participants were not familiar with the subject REST APIs or \evo-generated test cases, we complemented the surveys with an industrial study involving professionals who regularly use \evo.
 However, such an industrial study is primarily from one enterprise's perspective (i.e., Volkswagen AG), and other enterprises and programs might focus on different aspects.
Finally, our evaluation focused exclusively on \RESTAssured test cases; therefore, the results may not generalize to other programming languages, frameworks, or problem domains.

\textbf{Conclusion validity.}
To reduce threats related to statistical inference, we applied appropriate non-parametric tests, including Wilcoxon-Mann-Whitney U-tests for pairwise comparisons and Friedman tests for multiple treatments.
Surveys collected $39$ and $31$ responses, providing enough data to make meaningful descriptive and inferential analysis.

\section{Conclusions}
\label{sec:conclusions}

In this work, we proposed novel approaches for generating descriptive names  for automatically generated system tests for REST APIs.
Our empirical evaluation compared default (i.e., \numbered), rule-based, and LLM-based naming techniques through two surveys and a developer questionnaire within an industrial study at Volkswagen AG.
For these experiments, we used the state-of-the-art fuzzer \evo.

The results indicate that among rule-based approaches, \evoconditionshort consistently outperforms alternatives such as \evosimpleshort and \evoqueryshort in participant ratings, achieving the highest proportion of positive feedback and statistically significant effect sizes.
When compared to state-of-the-art LLM-based approaches (\emph{Gemini}, \emph{GPT-4o}, and \GitHub Copilot), \evoconditionshort performs on par, with no statistically significant differences observed, while outperforming \emph{GPT-3.5}.

Feedback from professional developers who regularly use \evo confirmed that \evoconditionshort provides clearer, more informative test names than the previous default numbering scheme.
Participants highlighted that some names could be further improved by including semantic information or more declarative suffixes.

Overall, our findings support the adoption of \evoconditionshort as the preferred naming strategy for automatically generated tests in \evo.
This approach combines high clarity, deterministic outputs, and reduced environmental and security risks, while avoiding the costs and risks of LLM-based approaches.

Future work could explore extending these techniques to other programming languages, frameworks, or types of system tests, and further refining rule-based naming to incorporate richer semantic annotations from REST API specifications.

\section*{Acknowledgments}

This work is funded by the European Research Council (ERC) under the European Union’s Horizon 2020 research and innovation programme (EAST project, grant agreement No. 864972), and partially funded by UBACYT-2020 20020190100233BA, PICT-2019-01793.


\bibliographystyle{ACM-Reference-Format} 

%

\bibliography{papers_arxiv}


\end{document}